\documentclass[10pt]{aa}  
\usepackage{comment}
\usepackage{orcidlink}

\hypersetup{linkcolor=blue,citecolor=blue,filecolor=black,urlcolor=blue}

\usepackage{amssymb}
\usepackage{ulem}

\usepackage{xspace}
\usepackage{pdflscape}

\usepackage{graphicx}

\usepackage[switch]{lineno} 
\usepackage{txfonts}
\begin{document}
%
%

   \title{ZTF SN~Ia DR2: Simulations and volume limited sample}
   \author{M.~Amenouche\inst{\ref{nrc}}\thanks{melissa.amenouche@nrc-cnrc.gc.ca}\orcidlink{0009-0006-7454-3579} 
    \and M.~Smith\inst{\ref{lancaster}}\thanks{mat.smith@lancaster.ac.uk}\orcidlink{0000-0002-3321-1432}  \and
	P.~Rosnet\inst{\ref{lpc}}\thanks{philippe.rosnet@clermont.in2p3.fr}\orcidlink{0000-0002-6099-7565}  
    \and
    M.~Rigault\inst{\ref{ip2i}}\orcidlink{0000-0002-8121-2560}  
    \and
    M.~Aubert\inst{\ref{lpc}}
    \and
    C.~Barjou-Delayre\inst{\ref{lpc}}
    \and
    U.~Burgaz\inst{\ref{trinity}}\orcidlink{0000-0003-0126-3999} 
    \and
    B.~Carreres\inst{\ref{marseille}, \ref{duke}}\orcidlink{0000-0002-7234-844X} 
    \and G.~Dimitriadis\inst{\ref{trinity}}\orcidlink{0000-0001-9494-179X} 
    \and F.~Feinstein\inst{\ref{marseille}}\orcidlink{0000-0001-5548-3466}
    \and
    L.~Galbany\inst{\ref{spain1}, \ref{spain2}}\orcidlink{0000-0002-1296-6887}
    \and
    M.~Ginolin\inst{\ref{ip2i}}\orcidlink{0009-0004-5311-9301}
    \and
    A.~Goobar\inst{\ref{albanova}}\orcidlink{0000-0002-4163-4996} 
    \and
    L.~Harvey\inst{\ref{trinity}}\orcidlink{0000-0003-3393-9383} 
    \and
     Y.-L.~Kim\inst{\ref{lancaster}}\orcidlink{0000-0002-1031-0796}
    \and
    K.~Maguire\inst{\ref{trinity}}\orcidlink{0000-0002-9770-3508}
    \and
    T.E.~M\"uller-Bravo\inst{\ref{spain1}, \ref{spain2}}
    \orcidlink{0000-0003-3939-7167} 
    \and
    J.~Nordin\inst{\ref{berlin}} \orcidlink{0000-0001-8342-6274} 
    \and
    P.~Nugent\inst{\ref{berkeley1}, \ref{berkeley2}}\orcidlink{0000-0002-3389-0586} 
    \and
    B.~Racine\inst{\ref{marseille}} \orcidlink{0000-0001-8861-3052}
    \and
    D.~Rosselli\inst{\ref{marseille}} \orcidlink{0000-0001-6839-1421}
    \and
    N.~Regnault\inst{\ref{lpnhe}} \orcidlink{0000-0001-7029-7901}
    \and
    J.~Sollerman\inst{\ref{albanova2}} \orcidlink{0000-0003-1546-6615}
    \and
    J.H.~Terwel\inst{\ref{trinity}}\orcidlink{0000-0001-9834-3439}
    \and
    A.~Townsend\inst{\ref{berlin}}\orcidlink{0000-0001-6343-3362}
    \and
    S.L.~Groom\inst{\ref{ipac}}\orcidlink{0000-0001-5668-3507}
    \and
S.R.~Kulkarni\inst{\ref{caltech}}\orcidlink{0000-0001-5390-8563}
    \and
     M.~Kasliwal\inst{\ref{caltech}}\orcidlink{000-0002-5619-4938}
    \and
    R.R.~Laher\inst{\ref{ipac}}\orcidlink{0000-0003-2451-5482}
    \and J.~Purdum\inst{\ref{caltech_obs}}\orcidlink{0000-0003-1227-3738}
    }

    \institute{
    National Research Council of Canada, Herzberg Astronomy \& Astrophysics Research Centre, 5071 West Saanich Road, Victoria, BC V9E 2E7, Canada\label{nrc} \and
    Department of Physics, Lancaster University, Lancaster, UK, LA14YW\label{lancaster} \and
     Universit\'e Clermont Auvergne, CNRS/IN2P3, LPCA, F-63000 Clermont-Ferrand, France \label{lpc} \and
     Univ Lyon, Univ Claude Bernard Lyon 1, CNRS, IP2I Lyon / IN2P3, UMR 5822, F-69622 Villeurbanne, France \label{ip2i} \and
    School of Physics, Trinity College Dublin, College Green, Dublin 2, Ireland \label{trinity} \and
    Aix Marseille Université, CNRS/IN2P3, CPPM, Marseille, France \label{marseille} \and
    Department of Physics, Duke University Durham, NC 27708, USA \label{duke} \and
    Institute of Space Sciences (ICE-CSIC), Campus UAB, Carrer de Can Magrans, s/n, E-08193 Barcelona, Spain \label{spain1} \and
    Institut d'Estudis Espacials de Catalunya (IEEC), 08860 Castelldefels (Barcelona), Spain \label{spain2} 
    \and
    The Oskar Klein Centre, Department of Physics, AlbaNova, Stockholm University, SE-106 91 Stockholm, Sweden \label{albanova} \and
    Institut für Physik, Humboldt-Universität zu Berlin, Newtonstr. 15, 12489 Berlin, Germany \label{berlin} 
    \and
    The Oskar Klein Centre, Department of Astronomy, AlbaNova, Stockholm University, SE-106 91 Stockholm, Sweden \label{albanova2}
    Lawrence Berkeley National Laboratory, 1 Cyclotron Road MS 50B-4206, Berkeley, CA, 94720, USA \label{berkeley1} \and
    Department of Astronomy, University of California, Berkeley, 501 Campbell Hall, Berkeley, CA 94720, USA \label{berkeley2} \and
    LPNHE, CNRS/IN2P3, Sorbonne Université, Université Paris-Cité, Laboratoire de Physique Nucléaire et de Hautes Énergies, 75005 Paris, France \label{lpnhe} \and
    Division of Physics, Mathematics, and Astronomy, California Institute of Technology, Pasadena, CA 91125, USA \label{caltech} \and
    IPAC, California Institute of Technology, 1200 E. California Blvd, Pasadena, CA 91125, USA \label{ipac}
    Caltech Optical Observatories, California Institute of Technology, Pasadena, CA 91125, USA \label{caltech_obs}
    }
   \date{}

\titlerunning{ZTF SN~Ia DR2 Simulation}
\authorrunning{M. Amenouche et al.}
 
  \abstract
  {Type Ia supernovae (SNe~Ia) constitute an historical probe to derive cosmological parameters through the fit of the Hubble-Lema\^{\i}tre diagram, i.e. SN~Ia distance modulus versus their redshift. In the era of precision cosmology, realistic simulation of SNe~Ia for any survey entering in an Hubble-Lema\^{\i}tre diagram is a key tool to address observational systematics, like Malmquist bias. As the distance modulus of SNe~Ia is derived from the fit of their light-curves, a robust simulation framework is required. In this paper, we present the performances of the simulation framework \texttt{skysurvey} to reproduce the the Zwicky Transient Facility (ZTF) SN~Ia DR2 covering the first phase of ZTF running from April 2018 up to December 2020. The ZTF SN~Ia DR2 sample correspond to almost 3000 classified SNe~Ia of cosmological quality. First, a  targeted simulation of the ZTF SN~Ia DR2 was carried on to check the validity of the framework after some fine tuning of the observing conditions and instrument performance. Then, a realistic simulation has been run using observing ZTF logs and ZTF SN~Ia DR2 selection criteria on simulated light-curves to demonstrate the ability of the simulation framework to match the ZTF SN~Ia DR2 sample. Furthermore a redshift dependency of SALT2 light-curve parameters (stretch and colour) was conducted to deduce a volume limited sample, i.e. an unbiased SNe~Ia sample, characterized with $z_{lim} \leq 0.06$. This volume limited sample of about 1000 SNe~Ia is unique to carry on new analysis on standardization procedure with a precision never reached (those analysis are presented in companion papers).}


   \maketitle


\section{Introduction}

In the past two decades, Type Ia supernovae (SNe Ia) have been used to probe the Universe on cosmological scales, typically up to a redshift $z \sim 1$.
Early measurements discovered the acceleration of the expansion of the Universe, while the recent state-of-the-art analysis allows to measure the mass density parameter ($\Omega_m$) within the $\Lambda$CDM model with an uncertainty better than two percent \citep{Brout:2022vxf}. This kind of analysis is based on the cosmological fit of the Hubble-Lema\^{\i}tre diagram, the SN Ia luminosity distance as a function of their redshift.

In current analysis, the SN Ia luminosity distance is determined from the fit of their light-curves in different photometric bands with a phenomenological model tuned on SN Ia Spectral Energy Distributions (SED). The most common used model is SALT2 \citep{SNLS:2007cqk}. But for precision cosmology, for instance to reach the two percent level precision in the cosmological parameters, the distance luminosity must be determined with a typical accuracy of 0.15~mag, i.e. with a photometric precision of typically 1\%.

Such precise measurements can be achieved only by correcting for any instrumental effects, but also for any systematic effect like the Malmquist bias, but not only as discussed in DES cosmological analysis paper \citep{vincenzi2024darkenergysurveysupernova}. The way to quantify any bias is to use realistic simulation, i.e. able to reproduce typical light-curves observed by the survey. With a robust simulation framework, high-statistic simulations can be generated to estimate any systematic effect starting from the SNe Ia light-curves that can affect the cosmological results.

At low redshift ($z < 0.1$) the DES Collaboration has shown that the bias on the SNe Ia distance modulus is limited to 0.05~mag, while at high redshift ($z > 0.1$) it can reach 0.4~mag \citep{DES:2018oqm}. To overcome the right distance modulus, in their simulation framework (SNANA software package, \citealt{snana_2009}), they include observational effects (sky-noise, zero-point...) and also detection efficiency determined from fake SNe Ia introduced in real images. This example illustrates how critical it is to accurately determine such bias for cosmological analysis.

The Zwicky Transient Facility (ZTF) is a low redshift survey \citep{Bellm_2019}. The necessity to measure with the best precision SN Ia peculiar velocity for bulk-flow analysis or for the measurement of $f\sigma_8$ is crucial \citep{Carreres_2023}. Beyond pure low redshift analysis, the combination of the unique ZTF SNe Ia sample with high redshift surveys, especially the coming one like LSST \citep{LSSTDarkEnergyScience:2022oih}, will also help in constraining better the cosmological parameters in the Hubble fit.

This paper highlights the tools developed by the ZTF Collaboration to simulate SN Ia light-curves in a realistic way by taking into account both the observing strategy and the instrumental limitations. In Sect.~\ref{sec-intrument}, the ZTF instrument is introduced as well as the ZTF Cosmology Data Release 2 of SNe Ia (ZTF SN~Ia DR2) corresponding to the three first years of observations. Then the simulation framework is described in section~\ref{sec-skysurvey}, while section~\ref{sec-framework} outlined the methodology developed to test the simulation pipeline. The test of the simulation framework based on the ZTF SN~Ia DR2 sample is discussed in section~\ref{sec-DR2}, before to conclude. 


\section{Data: The Zwicky Transient Facility}
\label{sec-intrument}

The Zwicky Transient Facility (ZTF) is a wide field optical survey covering the visible northern sky \citep{Bellm_2019}, using the Samuel Oschin 48-inch (1.2-m) Schmidt Telescope at the Palomar observatory \citep{Graham_2019}. The ZTF camera, consisting in a mosaic of 16 CCDs, has a field-of-view of $47\text{deg}^2$ \citep{Dekany_2020}, 13.3\%  of which is lost due to chip gaps and vignetting. The first phase of ZTF, called ZTF-1, with an average nightly cadence of every 2.5 nights in $g$ and $r$-band, and $5\sigma$ limiting depths of 20.5~mag, started observations in March 2018. All data obtained are processed, using difference imaging techniques by the Infrared Processing and Analysis Center (IPAC, \citealt{Masci_2019}) to estimate observing conditions (e.g. limiting magnitude, atmospheric distortion and zero-point) and detect candidates. This information is then disseminated to the community through alert packets \citep{Patterson_2019}. From March 2018 to December 2020, ZTF-1 discovered and classified 4156 supernov{\ae}.
A full description of the ZTF-1 observing pattern can be found in \textcolor{red}{Rigault et al. (a)}. 

\subsection{Spectroscopic Follow-up}

Spectroscopic follow-up for ZTF-1 was lead by the Bright Transient Survey (BTS, \citealt{Perley_2020,Fremling_2020}). This project aimed to produce a magnitude-limited census of extra-galactic transients by classifying all ZTF discoveries with $m_{\rm peak}<18.5$. Classifications were predominately obtained using the low resolution integral field unit, SEDm \citep{Blagorodnova_2018, Rigault_2019, Kim_2022} with additional time allocated on large facilities (see \citealt{Fremling_2020} for details) to classify missed or rapidly fading events. From ZTF-1, this project classified 3597 transients\footnote{See \href{https://sites.astro.caltech.edu/ztf/bts/explorer.php}{https://sites.astro.caltech.edu/ztf/bts/explorer.php} for latest values}, with completeness fractions estimated to 97\%, 93\% and 75\% for objects brighter than 18~mag, 18.5~mag and 19~mag, respectively (see \citealt{Perley_2020} for selection requirements and details).

\subsection{ZTF SN~Ia DR2}

ZTF-1 has discovered and classified 3628 SNe Ia. This second SNe Ia data release (ZTF SN~Ia DR2 detailed in \textcolor{red}{Rigault et al. (a)} and \textcolor{red}{Smith et al.}, called DR2 Throughout in the text), following ZTF SN~Ia DR1 presented in \cite{Dhawan_2022}, includes events discovered and classified from March 2018 to December 2020. About 80\% of events in this sample were classified by the BTS survey, with the remaining classifications obtained from public surveys (e.g. \citealt{Smartt_2015}) and individual spectra reported to the Transient Name Server (TNS\footnote{\href{https://www.wis-tns.org}{https://www.wis-tns.org}}). 
Photometry for this sample is determined using forced-photometry extracted from difference images, created, as part of the IPAC pipeline, by subtracting each science image from a stack of ZTF images taken during survey operations. A full description of the sample, and its photometric characterisation is given in \textcolor{red}{Smith et al.}, with companion papers on host galaxy properties and standardisation~\citep{ginolin2024ztf_1,ginolin2024ztf_2,popovic2024ztfsniadr2} with large-scale structure correlations \citep{ruppin2024ztfsniadr2, aubert2024ztfsniadr2}, light-curves properties \citep{rigault2024ztfsniadr2_2, deckers2024ztfsniadr2}, spectral properties \textcolor{red}{Johansson et al.} and diversity \citep{burgaz2024ztfsniadr2_1}, \textcolor{red}{Dimitriadis et al.} with astrophysics \textcolor{red}{Burgaz et al.} and cosmological \citep{carreres2024ztfsniadr2,dhawan2024ztfsniadr2cosmologyindependent} impacts. 

In this work, we aim to estimate the completeness for the DR2 in term of redshift defining a volume limited sample.


\section{Simulation Framework}
\label{sec-skysurvey}

To simulate the DR2 sample, we randomly generate samples of SNe Ia, drawn from a realistic model, each with a given redshift (drawn from a model describing the volumetric evolution of the rate of SNe~Ia events), and sky position. Each simulated SN~Ia is then matched to the \textit{true} ZTF observing cadence and observing conditions to predict the flux and the associated uncertainty (from which we can deduce the signal-to-noise ratio) that ZTF would have observed at each epoch, and produce simulated light-curves. By applying the selection criteria (accounting for photometric detection, light-curve characterisation and spectroscopic selection) used to define the DR2 sample, we predict the distribution of events in the DR2 sample.  

\begin{figure*}[!htb]
  \centering
  \includegraphics[width=\linewidth]{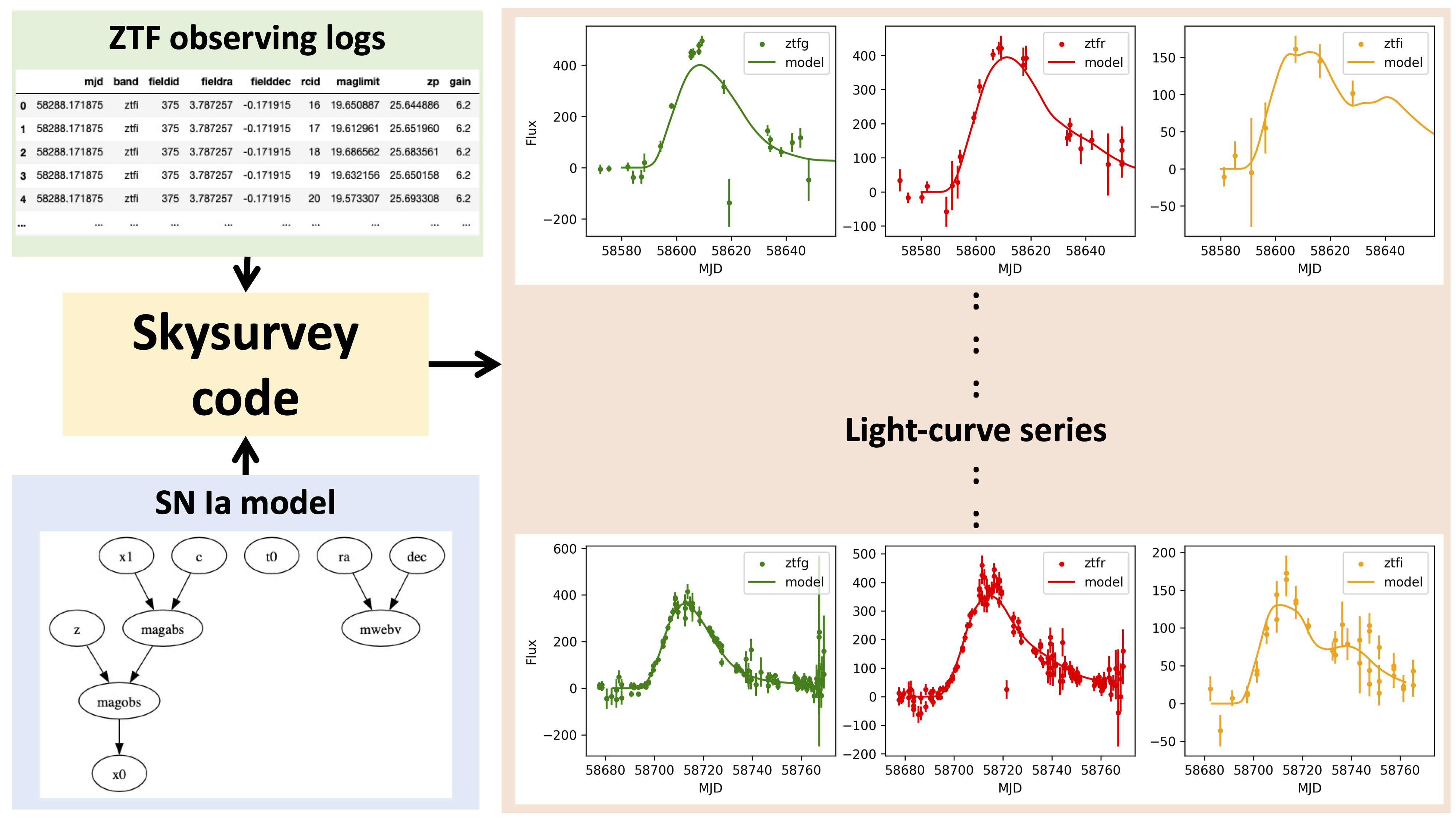}
  \caption{Schematic representation of our simulation pipeline. \emph{Left}: model light-curves are generated and combined with the true ZTF observing conditions to produce simulated light-curves. Selection criteria, matching those used to generate the ZTF SN~Ia DR2 sample are then applied to produce a mock ZTF SN~Ia DR2 dataset. \emph{Right}: Comparing our predicted dataset to the ZTF SN~Ia DR2 sample, we can estimate completeness and biases.
  } 
  \label{fig:schematic}
\end{figure*}

To do this we use a new python package \texttt{skysurvey}\footnote{\href{https://github.com/MickaelRigault/skysurvey}{https://github.com/MickaelRigault/skysurvey}}~\citep{skysurvey} developed for any transient survey, but described here for the purpose of ZTF. This new simulation framework is an improved (from computing point-of-view) version of \texttt{simsurvey}~\citep{Feindt_2019} build in a modular way (\texttt{simsurvey} has already been applied to model the ZTF survey \citealt{De_2020,SaguesCarracedo_2021,Magee_2022}). It allows to simulate thousands of astronomical targets as they would be observed by a survey. The main scheme is the same as for \texttt{simsurvey}: a targeted transient described by its model convoluted with an instrument cadence characterised by its sky deepness and its transmission filters produce light-curves. A schematic representation of this workflow is given in Fig.~\ref{fig:schematic}. The main characteristics of \texttt{skysurvey} are: (i) the coding which fasten the transient generation by about a factor thousand compared to \texttt{simsurvey} and (ii) a modular setup allowing to build any transient model by taking into account all astrophysical inputs as the environmental effects. This approach is consistent with the approach taken by \texttt{SNANA} \citep{snana_2009} framework commonly used in most SN cosmology analyses. 

This new framework was used to study the DR2 selection and its potential biases. To this end \texttt{skysurvey} can be used to simulate the typical SNe~Ia sample observed by ZTF during its first phase by using its observed log file, i.e. between $2018$ $1^{st}$ April to $2020$ December $31^{st}$ in g, r and i-bands. 

\subsection{SN Ia model}

To simulate an individual object, we first draw a redshift and sky position. To do this, we randomly pick-up a redshift from a non-evolving volumetric rate in the range $0 < z < 0.2$, based on Planck 2018~\citep{Planck_2018} fiducial cosmology (flat $\Lambda$CDM with $H_0=67.66~\text{km}\,\text{s}^{-1}\,\text{Mpc}^{-1}$ and $\Omega_m=0.30966$), and a sky position (RA, Dec) within the ZTF footprint~\citep{Dekany_2020}.
\texttt{Skysurvey} can be run in two way: (i) by using a defined rate, by the default  SN~Ia rate is  $2.35 \times 10^{-4}~\text{Gpc}^{-3}~\text{yr}^{-1}$~\citep{Perley_2020}, or (ii) by specifying a number of SNe Ia to simulate.

To generate a SN Ia light-curve, we use the SALT2 time-dependent spectral energy distribution model \citep{Guy_2007,Guy_2010}. This template parameterizes SNe Ia light curves by three parameters: $t_0$, the epoch of maximum brightness in $B$-band, $x_1$, the light-curve width, and $c$, the rest-frame colour. To ensure that our light-curves are representative of the overall SNe~Ia population, for each event we draw $x_1$ from the distribution defined in \cite{Nicolas_2021} using the parameters determined in \cite{ginolin2024ztf_1} and $c$ from that of \cite{Scolnic_2016}. $t_0$ is drawn randomly within the ZTF-1 observing time range. To set the apparent peak luminosity in $B$-band, $m_B$, we generate a peak absolute magnitude of $-19.3$, scattered using a colour independent  intrinsic dispersion of $\sigma_{\rm int}= 0.1$-mag and correct for the luminosity-decline rate and luminosity-colour relations with the nuisance parameters $\alpha=0.14$ and $\beta=3.15$ from \citep{Betoule_2014}, via a Tripp relation~\citep{Tripp_1998}. For each simulated event, we use the \texttt{sncosmo} framework \citep{Barbary_2016}, to generate observer-frame light-curves given the above information, and include the effect of the Milky Way (MW) dust by using the extinction maps from \cite{Schlegel_1998}. 

\subsection{Observing Logs}

To match our generated model with the ZTF-1 survey, we compiled the observing conditions for every ZTF observation by querying the IPAC database. In particular, for every observation, we retrieve\footnote{See \cite{Masci_2017} for a description of all columns available.}: 
\begin{itemize}
    \item date: \textit{mjdobs},
    \item sky location (observed field within ZTF grid): \textit{fieldid}, 
    \item ZTF-filter, CCD and readout-channel: \textit{fid}, \textit{ccdid} and \textit{qid},
    \item processing status: \textit{infobits},
    \item zero-point magnitude of the observation: \textit{magzp},
    \item $5~\sigma$ limiting magnitudes of the science image: \textit{scimaglim},
    \item readout channel gain: $gain$.
\end{itemize}

We convert the measured limiting magnitude to an effective sky brightness (hereafter referred to as skynoise) using 
\begin{equation}
s_b = \frac{10^{0.4 \times (magzp - scimaglim)}}{5}.
\label{sky-brightness}
\end{equation}

Our logs cover all observations from April, 2018 to December, 2020. A total of 431k exposures are retrieved, of which (37.1\%, 55.5\%, 7.4\%) are in (g, r, i)-band, respectively.

\subsection{Signal-to-Noise Ratio}
    
Flux uncertainties on each simulated event have three contributions: 
\begin{equation}
\sigma_{f} = \sqrt{s_b^{2} + \frac{|f|}{gain} + (f\times \sigma_{calib})^2},
\label{sigma_sn_calib}
\end{equation}
where $s_b$ is the sky brightness (defined in Eq.~\eqref{sky-brightness}), $f$ is the simulated flux, $gain \approx 6.2$ is the CCD gain and $\sigma_{calib}$ is the photometric calibration precision \citep{Masci_2019}.

Before applying this framework to free simulation, it was first tested on the DR2 targeted SNe Ia to check its validity to reproduce measured light-curves. In order to estimate the accuracy of the simulations, we compare the simulated quantities of the DR2 light-curves to the measured ones. The key quantity used for this purpose is the Signal-to-Noise Ratio (SNR) defined as following:
\begin{equation}
\text{SNR} = \frac{f}{\sigma_f},
\label{snr_formula}
\end{equation}
where $f$ and $\sigma_f$ are respectively fluxes and their associated uncertainties from Eq.~\eqref{sigma_sn_calib}. We compute the SNR of every DR2 object at each epoch of its measured and simulated light-curve for each band ($g$, $r$, $i$). 


\section{Testing the simulation framework}
\label{sec-framework}

In this section, we describe our work to validate the accuracy of \texttt{skysurvey} in reproducing the DR2 light-curves. By taking into account the real-time observing conditions and cadence of ZTF and the observed properties of the objects gathered from the data, we aim to replicate individual objects from the DR2 sample and compare the simulated and measured light-curves. In \ref{sub:metho}, we describe the methodology we followed, in \ref{sub:exp} we show the comparison of a simulated and measured light-curve for one object of the sample, then in \ref{sub:all-sample} we discuss the comparison of the fluxes, their uncertainties and their SNR for the whole sample.

\subsection{Methodology} 
\label{sub:metho}

To simulate the DR2 sample, we use the observed properties of the sample, \textit{i.e} the SALT2 \citep{Guy_2007, Guy_2010} fitted parameters of the measured light-curves as \texttt{skysurvey} inputs. The SALT2 parameters of the sample are presented in \cite{rigault2024ztfsniadr2_2}. We use $t_{0}$, $x_{1}$, $c$ and $z$ of every individual object in \texttt{skysurvey} combined to ZTF observing logs to simulate every SN~Ia. In order to evaluate the accuracy of our simulations, we compare the fluxes, flux uncertainties and SNR of the simulated and measured light-curves at all epochs for all DR2 objects. Furthermore, to avoid statistical fluctuations, each SN~Ia is simulated 10 times, meaning that for the whole DR2 sample, we generated about 30,000 SNe~Ia.

\subsection{One example}
\label{sub:exp}

\begin{figure*}[!htb]
  \centering
  \includegraphics[width=1.0\linewidth]{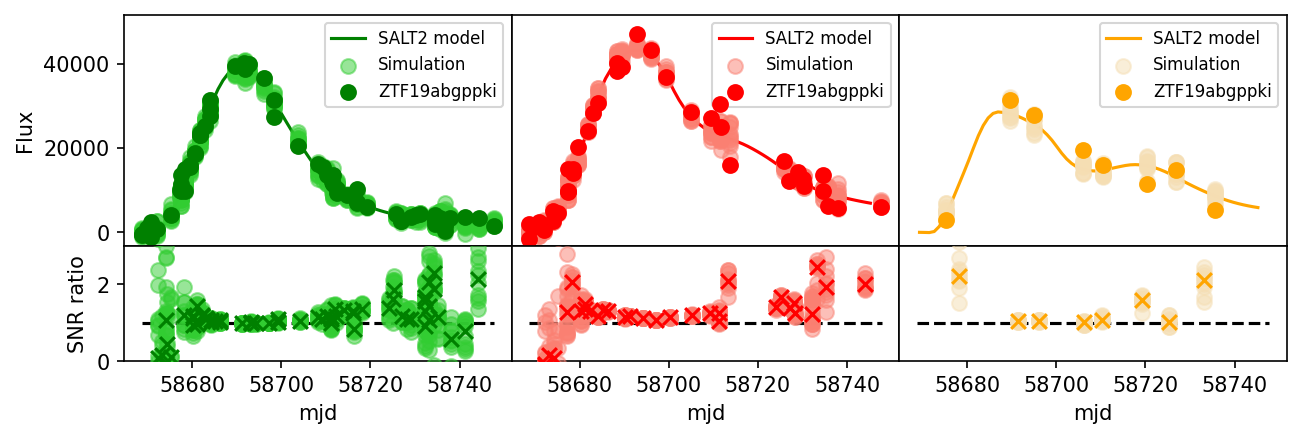}
  \caption{\textit{Top:} Simulated (10 times) and measured ZTF19abgppki ($z=0.06096$) light-curves in $g$ (left), $r$ (middle) and $i$ (right) bands. \textit{Bottom:} evolution of the SNR ratio of the simulated light-curves w.r.t. measured data points as a function of time (in $mjd$). The crosses show the median of the 10 simulations for the SNR ratio for each observation date.  
  } 
  \label{fig:ex_lc}
\end{figure*}

As an example, Fig.~\ref{fig:ex_lc} shows the measured and simulated light-curve of ZTF19abgppki, in $g$ (left), $r$ (middle) and $i$-band (right). The top part of each plot shows a good agreement between the simulated and measured fluxes in every three bands at all epochs. We compute the SNR values of the simulated and measured light-curves and their ratios. The bottom part of each plot in Fig.~\ref{fig:ex_lc} shows the evolution of the SNR ratios (SNR$_{sim}$/ SNR$_{data}$) as a function of time (in $mjd$) for the three bands. The crosses represent the median SNR values per observation. We can observe that the SNR ratios stand above $1$, especially for low-fluxes at the starting point and in the tail of the light-curves, indicating that the simulated light-curves display higher SNR than the DR2 ones. And as the fluxes seem to match in the upper part of the plots of Fig.~\ref{fig:ex_lc}, it indicates that the simulated uncertainties are not reproducing the measured ones. To get a quantitative idea of the discrepancy, and then the correction to apply in our simulations, a statistical analysis over the full DR2 objects is necessary.

\subsection{Statistical test of \texttt{skysurvey}}
\label{sub:all-sample}

From the 3628 SNe~Ia of the full DR2 sample, we apply the following set of basic cuts on the DR2 light-curves properties from \textcolor{red}{Rigault et al. (a)}
\begin{itemize}
    \item $|x_{1}| < 3$,
    \item $c \in [-0.2, 0.8]$,
    \item $\delta t_0 \leq 1$,
    \item $\delta x_{1} \leq 1$,
    \item $\delta c \leq 0.1$.
    \item probfit $> 10^{-7}$,
    \item we require data points with a phase ($\Phi$) $\in [-10, 40]$.
\end{itemize}
We apply additional cuts of good sampling, they can be found in \textcolor{red}{Rigault et al. (a)}. The phase cut is dictated by the study of DR2 light-curve residuals led by \cite{rigault2024ztfsniadr2_2}. The final sample counts $2662$ SNe~Ia. For the purpose of our study, we will use this sample as our baseline.

So, we simulate 10 times those 2662 SNe~Ia with their SALT2 parameters and their redshift as inputs to \texttt{skysurvey}. 

\subsubsection{Light-curves comparison}
\label{sub:light-cuves}

To ensure an accurate comparison between the measured and simulated quantities, for every SN~Ia we match its measured and simulated light-curve along with the associated observing logs. 

\begin{figure}[hbt!]
  \centering
  \includegraphics[width=\linewidth]{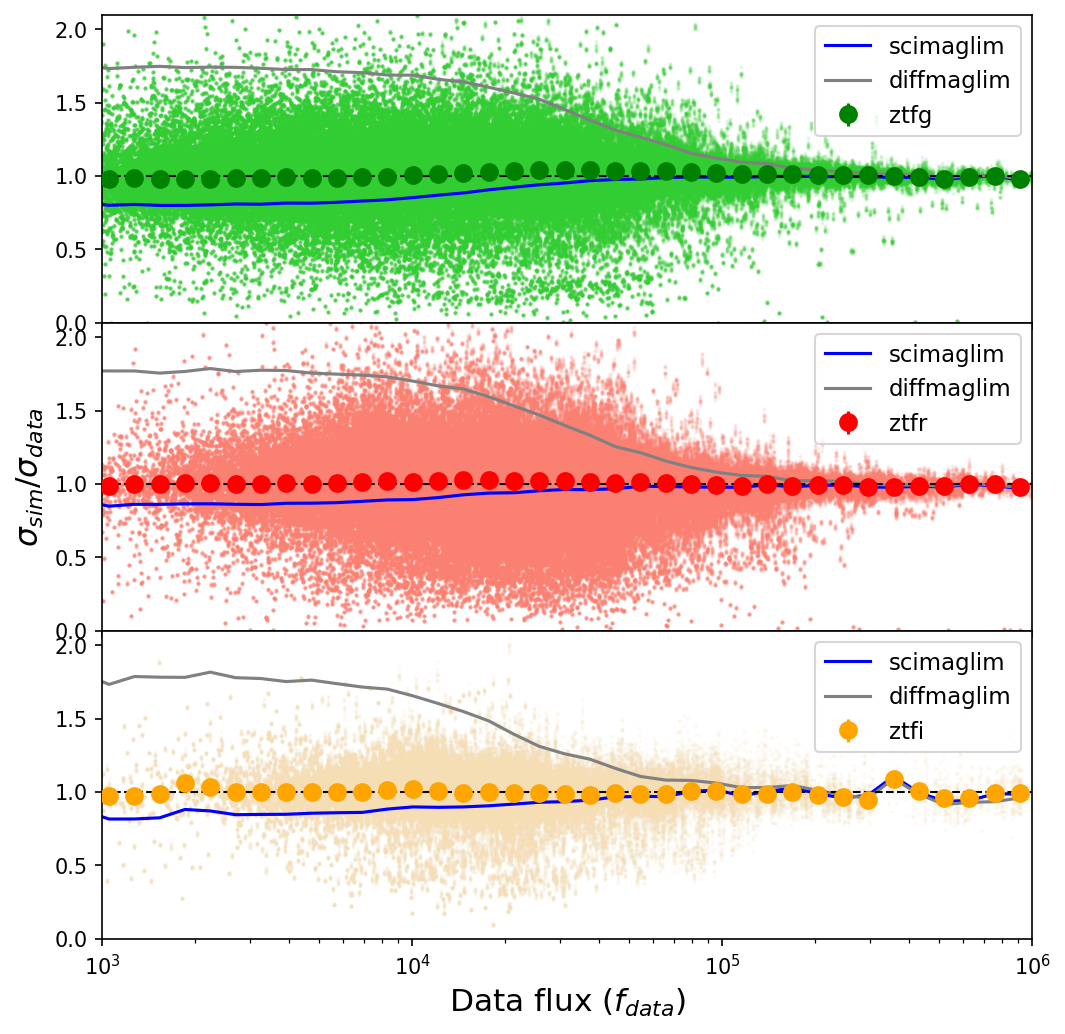}
  \caption{Ratio of simulated and DR2 flux uncertainties ($\sigma_{sim}/\sigma_{data}$) as a function of data flux in $g$ (\textit{top}), $r$ (\textit{middle}) and $i$ (\textit{bottom}) bands. Small points are comparison between every simulated and observed objects at every phase, while the bigger points are the associated median binned values, both after skynoise correction (see text for details). The blue curves ($scimaglim$) are the binned values obtained from the science-image, while the grey curves ($diffmaglim$) are the binned values obtained from the difference-image limiting magnitudes.}
  \label{fig:error_comparison_g_r}
\end{figure}

The DR2 fluxes and associated uncertainties are extracted from the difference images, based on reference images, as discussed in the \textcolor{red}{Smith et al.} companion paper. So, we computed the flux uncertainties following Eq.~\eqref{sigma_sn_calib} using skynoise derived from the difference image limiting magnitudes or from the science image limiting magnitudes. The ratio of the simulated uncertainties and the DR2 associated ones ($\sigma_{sim}/\sigma_{data}$) are shown in Fig.~\ref{fig:error_comparison_g_r} for $g$ (top), $r$ (middle) and $i$ (bottom) bands: as grey curves for the difference image limiting magnitudes ($diffmaglim$) and as blue curves for the science image limiting magnitudes ($scimaglim$).

    
At low flux (typically $f < 10^4$), where the skynoise contribution (Eq.~\eqref{sigma_sn_calib}) is dominant, we can see that the ratios of the simulated uncertainties computed with skynoise from the difference image limiting magnitude and the DR2 one are systematically above $1$, while the ratios of the uncertainties computed from the science limiting magnitude are systematically below $1$. Furthermore, departure from $1$ is more pronounced with the difference image limiting magnitude, that the limiting magnitude associated to the difference image is not correctly estimated. In the case where the skynoise is estimated using the science image limiting magnitude, the simulated flux uncertainties are under estimated compared to the measured ones, which is expected because the difference image processing add some background noise, but the resulting values are closer. It is a clear indication for the need of a new estimated skynoise in the simulations, in order to accurately replicate the DR2 light-curves. To estimate the new skynoise, we use data points which are at low flux ($ f < 5000$). We compute the median of the measured and simulated flux uncertainties per flux ranges. By computing the median of the binned measured and simulated uncertainties we obtain the corrective factor for the skynoise. We find that a corrective factor of $1.23$ for g-band, $1.17$ for r-band and $1.20$ for i-band are needed to solve the discrepancy between the simulated and measured uncertainties. This means that we need to increase the skynoise level associated to science image to account for the forced photometry output which is applied to the difference images. To obtain the values of the new skynoise, we multiply the skynoise obtained from the science limiting magnitude by the corrective factors we estimated for each of both bands. 
 
In addition, an error-floor uncertainty of $2.5\%$, $3.5\%$ and $6\%$ flux level, in respectively $g$, $r$ and $i$-bands, is added to the DR2 flux uncertainties \cite{rigault2024ztfsniadr2_2}. In order to accurately compare the DR2 flux uncertainties to the simulated ones, we need to account for this error-floor in the simulation framework. 
    
We simulate back our sample with the correct prescription for the skynoise and the error-floor. The ratio of the newly simulated uncertainties and the measured ones for the sample are shown in the same figure (Fig~\ref{fig:error_comparison_g_r}). We compute the median values for the ratios per flux range, they are represented in the same figure by the bigger points. We can notice the data points are scattered around $1$, with a bigger scatter at low-flux due to the skynoise dominance in the flux uncertainties. We can also notice that the binned data points are between the ratio of uncertainties computed with the science and difference images limiting magnitudes. The binned ones are compatible with $1$ along with their evolution with the flux, indicating that the newly simulated flux uncertainties and the measured ones are in good agreement. 

\begin{figure}
    \includegraphics[width=\linewidth]{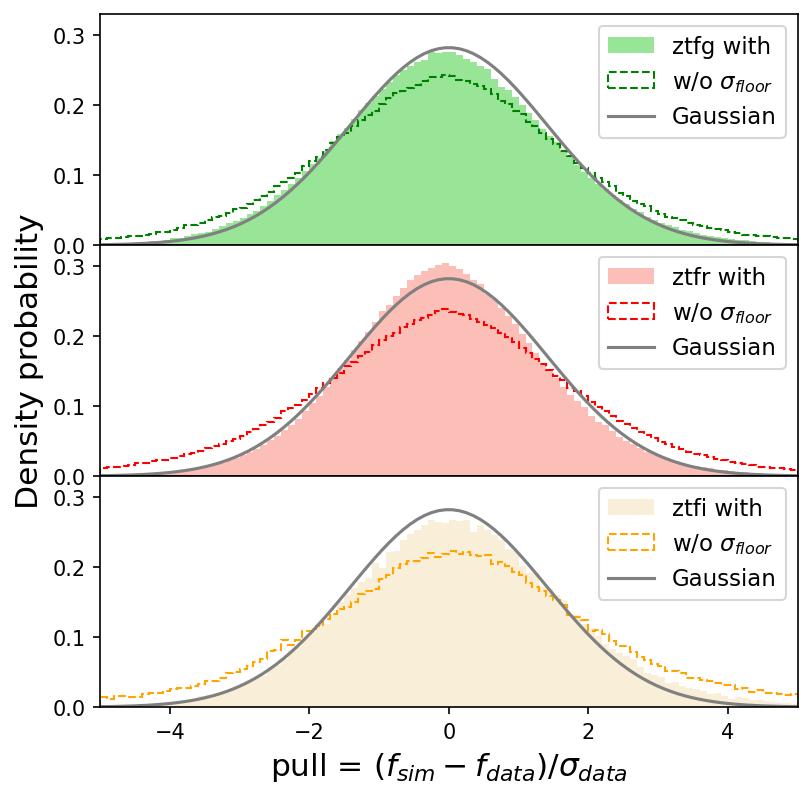}
    \caption{Light-curve pulls = (flux$_{sim}$ - flux$_{data}$) / $\sigma_{data}$ for $g$ (\textit{top}), $r$ (\textit{middle}) and $i$ (\textit{bottom}) bands. The full (open dashed) histograms show the pulls with (without) error-floor, while the grey curves correspond to a Gaussian centered on zero and with a sigma equal to $\sqrt{2}$.}
    \label{fig:pulls}
\end{figure}

As discussed in~\cite{rigault2024ztfsniadr2_2}, an error-floor is necessary to account for the Gaussian distribution of light-curve data points w.r.t. SALT2 model along the full fitted phase range [-10, 40]. The simulated fluxes were compared to the measured ones by computing the pull $= (f_{\it sim} - f_{\it data}) / \sigma_{\it data}$ for every points of the 10 simulated light-curves of the full DR2. Fig.~\ref{fig:pulls} shows the pull distribution for the three bands. The full histograms correspond to the pull computed by adding the error-floor, while the open dashed histograms are the pull computed without the error-floor. As light-curve data points are distributed following a 1-sigma Gaussian centered on zero w.r.t. their SALT2 fitted model (see Fig.~1 of~\cite{rigault2024ztfsniadr2_2} where $\sigma_{\it data-pull} = 1$) and as the simulated light-curves are generated following a random Gaussian distribution w.r.t. to the SALT2 model (i.e. such as $\sigma_{\it sim-pull} = 1$), we expect for the pull of Fig.~\ref{fig:pulls} a Gaussian distribution centered on zero with a sigma equal to $\sigma_{\it pull} = \sqrt{\sigma_{\it data-pull}^2 + \sigma_{\it sim-pull}^2} = \sqrt{2}$. This prediction is represented by the solid grey curves on Fig.~\ref{fig:pulls}. The comparison between the expected pull and the computed one are in good agreement when taking into account the error-floor (full histograms), while the computed pull distributions are wider when the error-floor is omitted. This result shows that the amount of error-floor of $2.5\%$, $3.5\%$ and $6\%$, respectively in $g$, $r$, and $i$-bands, are necessary for a good matching between simulations and the DR2 data.
   
\subsubsection{Signal-to-Noise Ratio comparison}
\label{subsub:snr}

\begin{figure}[hbt!]
  \centering
  \includegraphics[width=\linewidth]{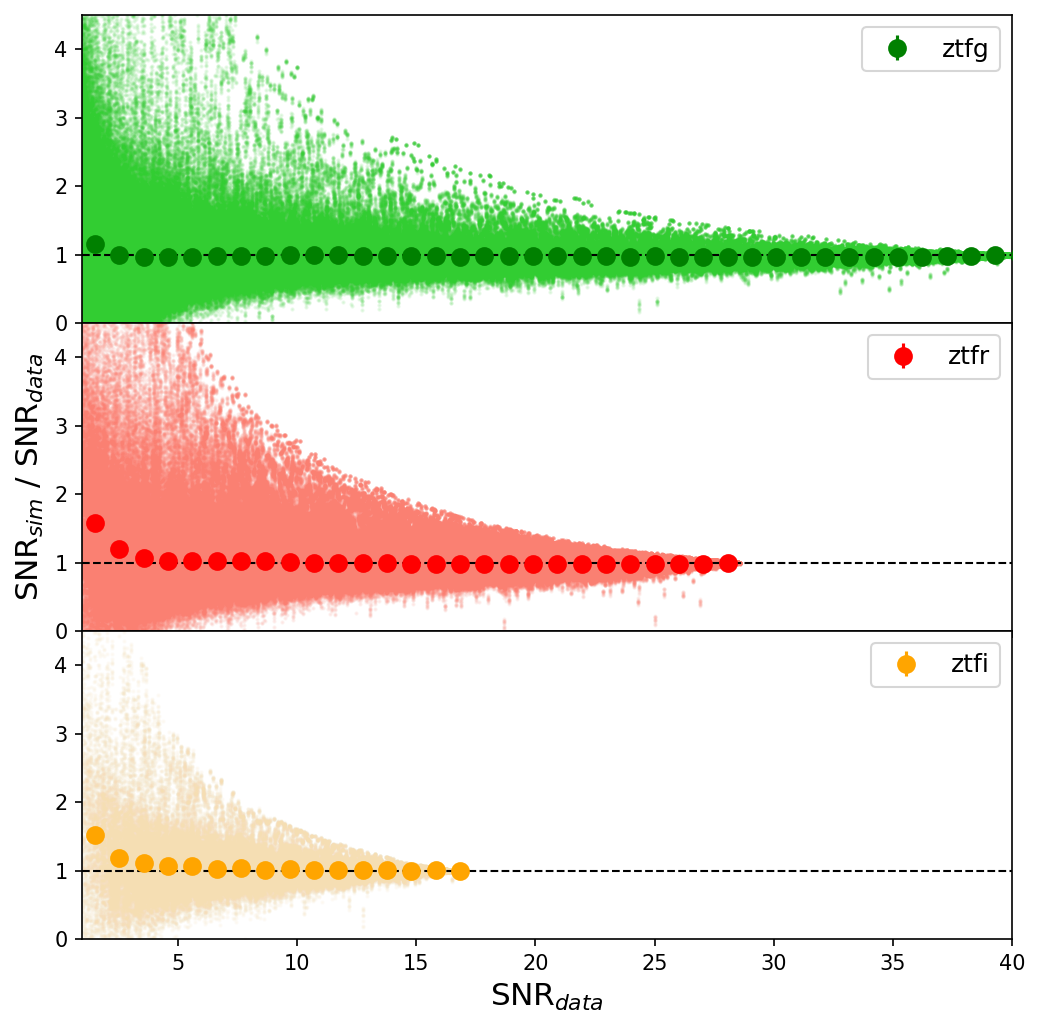}
  \caption{SNR ratio of simulated and DR2 as a function of DR2 SNR in $g$ (\textit{top}), $r$ (\textit{middle}) and $i$ (\textit{bottom}) bands. Small points are comparison between every simulated and observed objects at every phase, the bigger points are the associated median binned values.} 
  \label{fig:snr}
\end{figure}

We use the newly simulated light-curves, with the prescriptions, from \ref{sub:light-cuves}, for the skynoise computed from the science limiting magnitudes with the correcting factors and the error-floor needed in the flux uncertainties. We compute the SNR of the simulated and observed data points at all epochs, following Eq~\eqref{snr_formula}. We represent the ratio of the SNR from the simulation and the DR2 as a function of the DR2 one in Fig.~\ref{fig:snr} for $g$ (top), $r$ (middle) and $i$ (bottom) bands. We compute the median of the SNR ratios per SNR data range, they are represented in the same figure with bigger points. One can notice that the data points display a wide scatter especially at low SNR, due to the skynoise fluctuations, knowing its dominance in the flux uncertainties budget at these SNR values. The scatter is reduced at higher SNR. The binned SNR ratios are consistent with $1$ in the three bands for SNR $> 5$. For the low-SNR ($< 5$), the data points are below the detection limit and the SNR ratio is affected by negative fluxes, so the SNR ratio ($sim / data$) $=1$ is no more valid. It shows that the simulations framework replicates the fluxes and associated uncertainties of the measured DR2 light-curves at all epochs, across ZTF redshift range.
  
\subsection{Summary}
\label{sub:summary_target_sim}

We aimed to simulate the most realistic DR2 SN~Ia sample, with all our knowledge of ZTF observations and SN~Ia modelling. We use the ZTF true observing strategy and state-of-the-art SN~Ia models in \texttt{skysurvey} to replicate the whole DR2 sample. We simulated $2662$ SNe~Ia, passing the good sampling and the basic cuts and tested multiple configurations of the simulation framework. We replicate the DR2 fluxes, associated uncertainties and SNR by using the skynoise computed from the science limiting magnitudes with a corrective factor of $1.23$ for g-band, $1.17$ for r-band and $1.20$ for the i-band. In addition, we account for an error-floor of $2.5\%$, $3.5\%$ and $6\%$ of the flux level in respectively $g$, $r$ and $i$-band. With these prescriptions, we can accurately simulate the ZTF DR2 SNIa light-curves at all epoch for every band ($g$, $r$, $i$) up to the redshift limit of the survey $z \approx 0.15$.


\section{Simulating the ZTF SN~Ia DR2 sample}
\label{sec-DR2}

Simulations are crucial for correcting biases in the analysis of SNe~Ia to infer cosmological quantities. In this section, we present our work on the DR2 selection sample with simulations. 

\subsection{ZTF SN~Ia DR2 simulation}

To study selection effects on DR2 sample a high-statistic simulation has been carried on. To avoid limited sample at very low redshift, this simulation was partitioned in redshift bins of 0.005 starting from 0 to 0.15, i.e. by generating 25k SNe~Ia in 30 redshift bins. Then a weighting procedure is applied to retrieve the natural power law redshift evolution.
SALT2 model is used for generation with a color parameters $c$ drawn from an asymmetric distribution following the model from \cite{Scolnic_2016} and a stretch parameters $x_1$ drawn from a bimodal distribution from~\cite{nn_21}.

All selection cuts applied to the DR2 sample are reproduced to the simulated light-curves. First, the BTS sampling cuts (see section 2.3 of~\citealt{Perley_2020}) with the selection function (see Fig.~4 of~\citealt{Perley_2020}) are applied to the generated light-curve data points. Second, the basic DR2 cuts (see table~2 of \textcolor{red}{Rigault et al. (a)}) are applied to the fitted light-curves.

\begin{figure}
  \centering
  \includegraphics[width=1.0\linewidth]{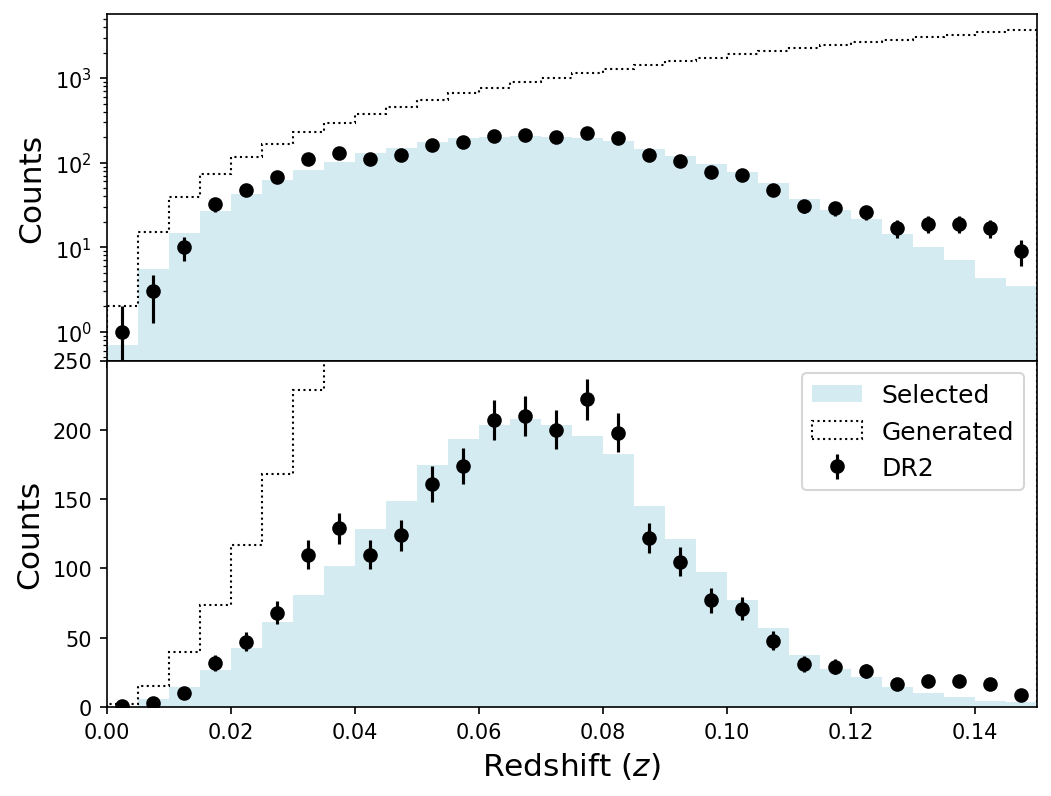}
  \caption{Comparison of DR2 redshift distribution (black points) with \texttt{skysurvey} simulation in log scale (upper plot) and linear scale (lower plot) from generated SNe~Ia distribution (black dotted line) to selected ones (full blue histograms). All simulated distributions (before and after selection cuts) are normalized to the data sample. 
  } 
  \label{fig:sim_z}
\end{figure}

All together, the selections cuts allow to reproduce the DR2 redshift distributions as shown in Fig.~\ref{fig:sim_z}.
The natural redshift power law is well decreased by all selection cuts to be able to reproduce DR2 redshift distribution. The matching is not perfect but in very good agreement. Indeed the high-redshift part ($z > 0.07$) reproduces the drastic reduction due mainly to the selection function up to a factor 100 at $z=0.15$. In the low-redshift regime, a small bump is seen in DR2 data compared to the smooth simulation especially in the range $0.03 < z < 0.04$. The origin of this bump has not been identified, but it might arise from the "local" large-scale structure of the Universe within the ZTF footprint or an unmodelled population of underluminous SNeIa.

\begin{figure}
  \centering
  \includegraphics[width=1.0\linewidth]{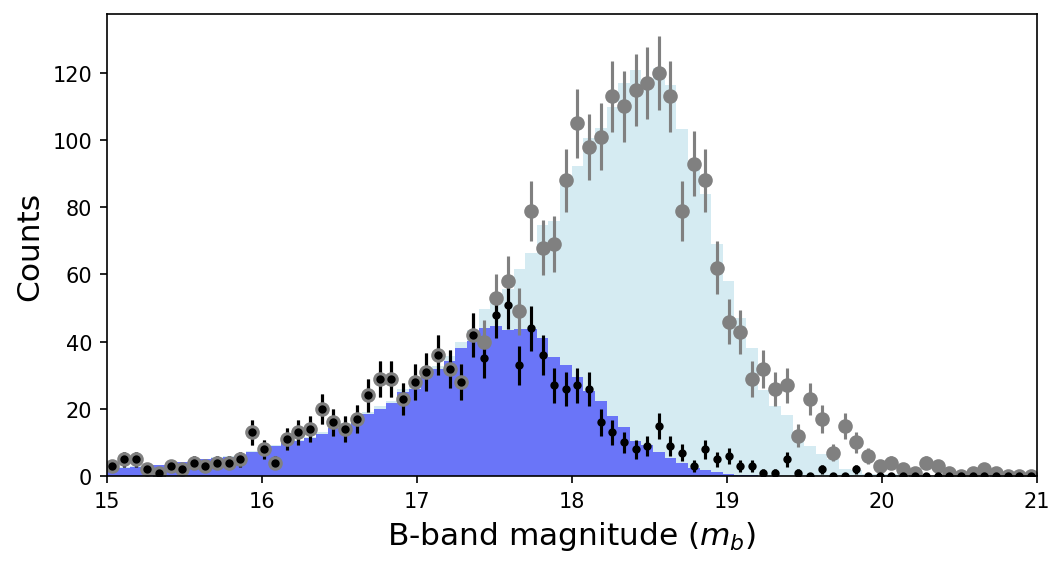}
  \hfill
  \includegraphics[width=1.0\linewidth]{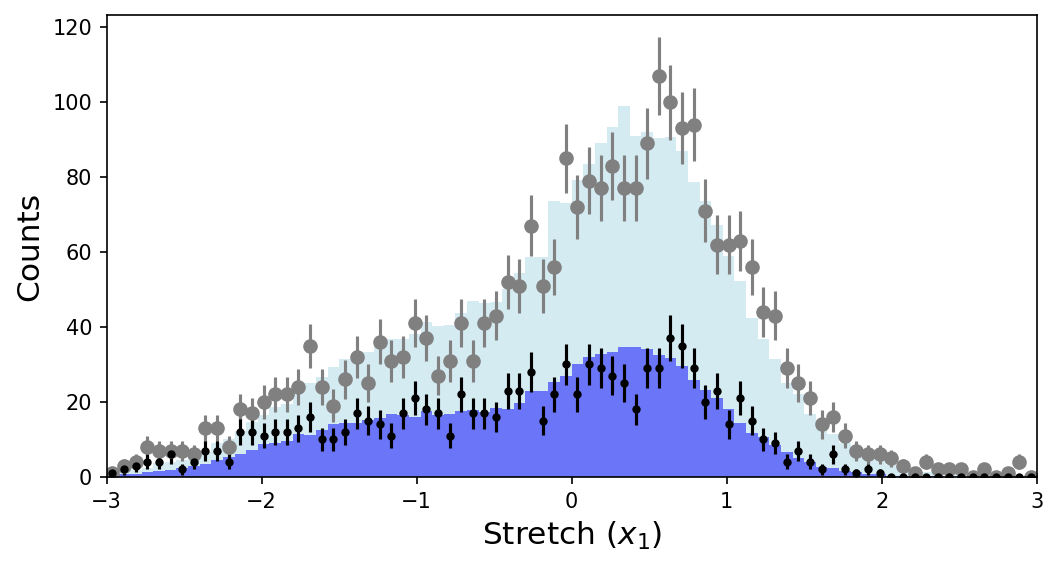}
  \hfill
  \includegraphics[width=1.0\linewidth]{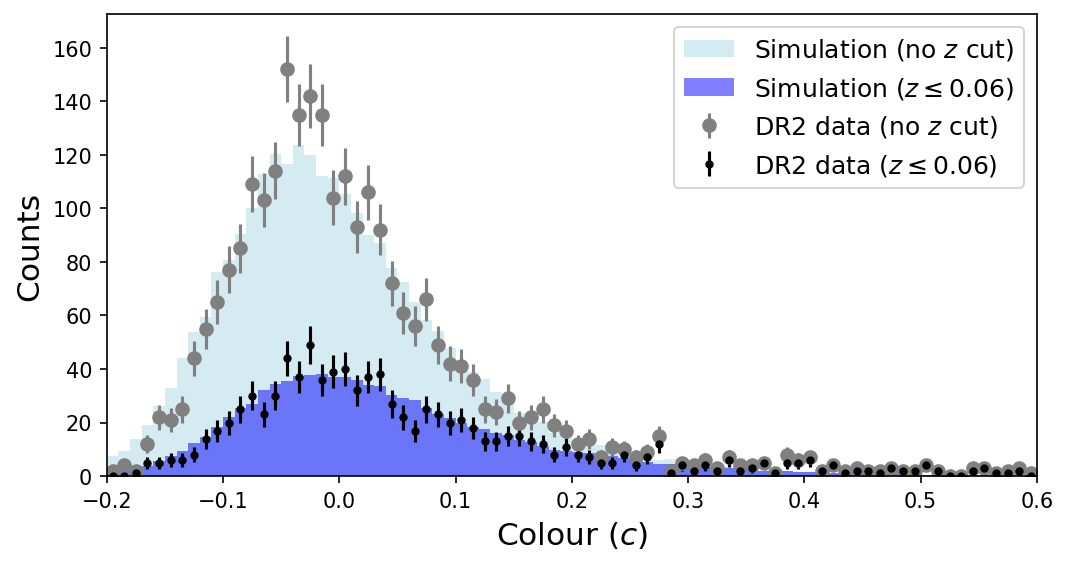}
  \caption{
  Comparison of DR2 SALT2 parameters (points with error bars) -- $m_b$ (upper plot), $x_1$ (middle plot) and $c$ (lower plot) -- with \texttt{skysurvey} SNe~Ia simulation after light-curves fit (histograms): without redshift cut (grey points and light-blue histogram) and for $z \leq 0.06$ (smaller black points and blue histogram).
  } 
  \label{fig:sim_salt2}
\end{figure}

Simulated light-curves fitting is conducted as for real data using the SALT2 model. A comparison of the main outputs from SALT2 parameters is shown in Fig.~\ref{fig:sim_salt2}: without redshift cut (grey data points and light-blue histogram for simulation) and for $z \leq 0.06$ (smaller black data points and blue histogram for simulation).
The three SALT2 output parameters are, at first order, well reproduced by the simulation, especially for the low-redshift selection $z \leq 0.06$.
We observed clearly the Malmquist bias for faint SNe~Ia in the $m_b = -2.5 \log10(x_0) + 10.635$ distribution (Fig.~\ref{fig:sim_salt2}, upper plot): compared to the low-redshift sample ($z \leq 0.06$), the full sample shows a steeper distribution at high $m_b$ illustrating the selection effect for higher redshift.
On the other side, the shape of the stretch ($x_1$, Fig.~\ref{fig:sim_salt2}, middle plot) and the colour ($c$, Fig.~\ref{fig:sim_salt2}, lower plot) distributions look the same (at first order) for the full sample and the low-redshift selection. A deeper analysis of the redshift dependence of these distribution is discussed in the next section.

To quantify the ability of the simulation to reproduce DR2 distributions, we computed a sort of chi-square per degree-of-freedom defined as 
\begin{equation}
\chi^2_{/ndf} = \frac{1}{N_{bin}} \sum_{i = 1}^{N_{bin}} \frac{(N_{data,i} - N_{sim,i})^2}{N_{data,i}}
\label{chi2_ndf}
\end{equation}
where $N_{bin}$ is the number of bins in the compared histograms which plays the role of degree-of-freedom. In this definition, we assume a Gaussian uncertainty for the data counting per bin ($\sigma_{data,i} = \sqrt{N_{data,i}}$) and we neglect the statistical fluctuation of the simulation (the number of simulated SNe~Ia passing all selection cuts is 47 times the DR2 dataset). The limits of the histograms are defined by cutting the $0.5\%$ of the DR2 left and right tails using the quantile method, i.e. keeping $99\%$ of the bulk of the distributions. For the robustness of the analysis, we varied $N_{bin}$ from 80 to 100 by step of 1 bin (i.e. 20 $\chi^2_{/ndf}$ calculations for each distribution comparison) and then we computed the mean $\chi^2_{/ndf}$ and evaluated an uncertainty on the mean value.

The results are presented in Table~\ref{tab:SALT2chi2ndf} for the three SALT2 parameters ($m_b$, $x_1$ and $c$) for the full DR2 sample (no $z$ cut) and the lower redshifts part ($z \leq 0.06$). For the full data sample, the chi-square per degree-of-freedom ($\chi^2_{/ndf}$) is between 1.4 and 1.7, depending on the SALT2 parameter, meaning that the simulations are not reproducing perfectly the full data-set. The worst agreement is observed for the B-band magnitude ($m_B$) parameter. When comparing DR2 and simulated distributions for the lower redshift part ($z \leq 0.06$, which corresponds to what we call the volume limited sample, see Sect.~\ref{sec-volume_limited_sample}) we observe a small improvement for the $m_B$ distribution, a much better matching for the stretch ($x_1$) distributions, and a consequent improvement for the colour distribution ($c$) for which the $\chi^2_{/ndf}$ is decreased by 25\%. This quantitative comparison shows that the simulation is close to reproduce the main properties of our SN~Ia data sample when considering the volume limited sample ($z \leq 0.06$) which is described in more detail in the next section.

\begin{table}
\centering
\caption{$\chi^2_{/ndf}$ (with an estimation of its uncertainty) between DR2 and simulation distributions of SALT2 parameters ($m_b$, $x_1$ and $c$) for the full sample (no $z$ cut) and for the volume limited sample ($z \leq 0.06$).}
\label{tab:SALT2chi2ndf}
\begin{tabular}{l c c c}
\hline
$\chi^2_{/ndf}$ & $m_b$ & $x_1$ & $c$ \\
\hline
no $z$ cut  & $1.62 \pm 0.03$ & $1.42 \pm 0.03$ & $1.64 \pm 0.03$ \\
$z \leq 0.06$ & $1.61 \pm 0.05$ & $1.26 \pm 0.03$ & $1.23 \pm 0.03$ \\
\hline
\end{tabular}
\end{table}

\subsection{Volume limited sample}
\label{sec-volume_limited_sample}

\begin{figure}
  \centering
  \includegraphics[width=1.0\linewidth]{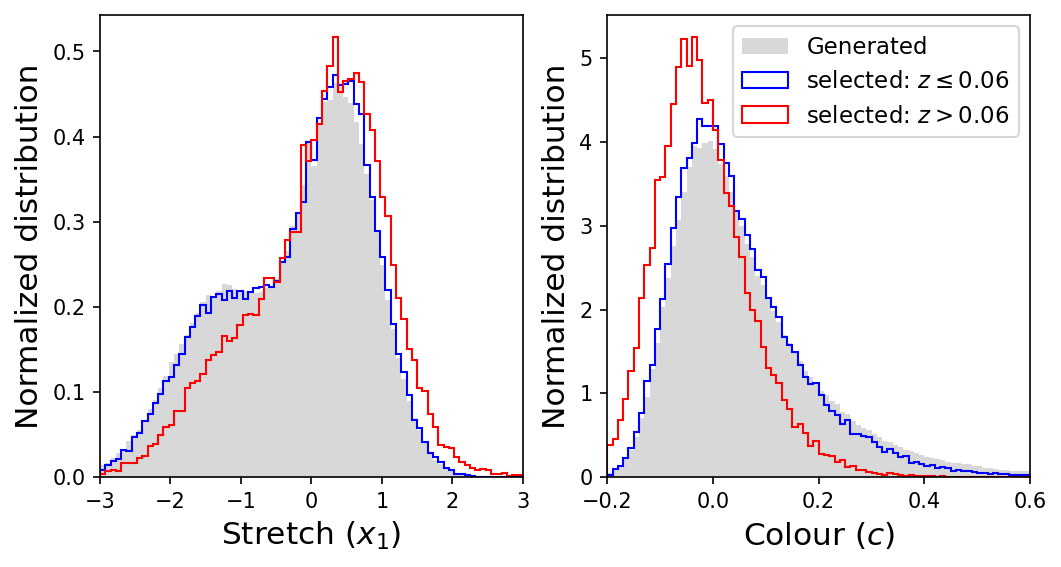}
  \caption{Comparison of stretch (left plot) and colour (right plot) input distributions before (grey filled histograms) and after all selection cuts (open histogram) for lower redshifts ($z \leq 0.06$, open blue histogram) and higher redshifts ($z > 0.06$, open red histogram).
  } 
  \label{fig:sim_x1c_z}
\end{figure}

An analysis of SALT2 parameters as a function of the redshift is necessary to estimate the limited redshift for which SNe~Ia are not biased. Fig.~\ref{fig:sim_x1c_z} shows a comparison between the input stretch (left plot) and colour (right plot) distributions before (grey filled histograms) and after selection cuts (open histograms), as for the ZTF-DR2-Cosmos data, for cumulative redshift ranges sampling the bulk of the full ZTF redshift interval. From those plots it appears that input distributions are well reproduced at very low redshift and start to be distorted when redshift increased beyond 0.05.

\begin{figure}
  \centering
  \includegraphics[width=1.0\linewidth]{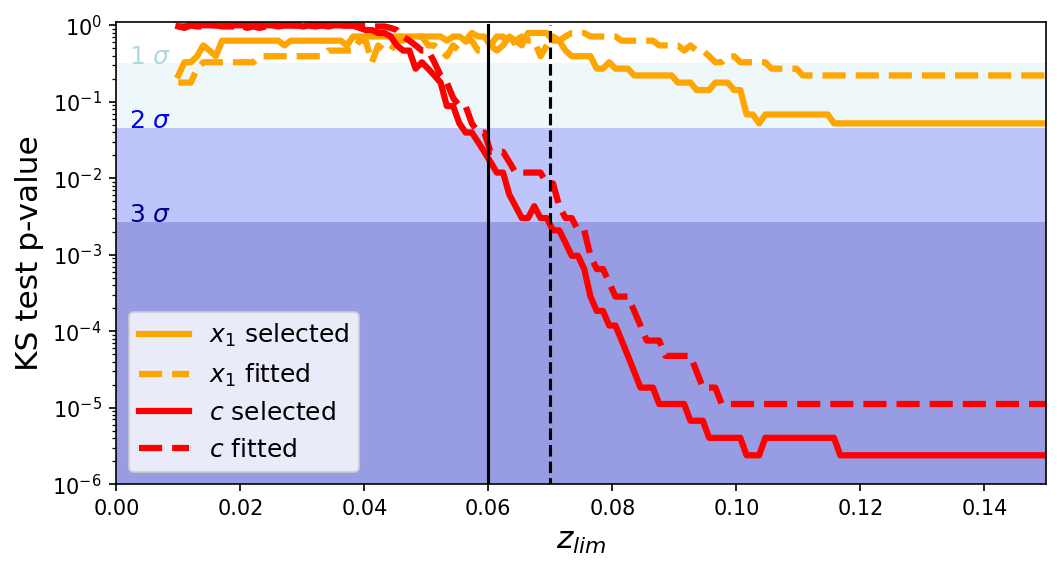}
  \caption{KS test p-value for stretch (orange curve) and colour (red curve) as a function of the redshift limit: full lines show the comparison between generated distributions after w.r.t. before selection, while dashed lines compare distributions of fitted SALT2 parameters after selection w.r.t. to generated distributions before selection. 
  } 
  \label{fig:sim_KS}
\end{figure}

A volume limited sample can be defined as a data sample for which the input SALT2 distributions are not significantly distorted, meaning that there is no bias in the selected SNe~Ia. To quantify the redshift limit of the volume limited sample a Kolmogorov-Smirnov (KS) test has been performed between the input distributions used for generation and the distributions retrieved after all selection cuts applied to simulated sample, as for real data, and as a function of redshift limit ($z < z_{lim}$). The result of the KS test of both stretch and colour parameters is shown in Fig.~\ref{fig:sim_KS} and compared to the probability limits at 1, 2 and 3~$\sigma$. This comparison allows to estimate the redshift limit $z_{lim} \approx 0.07$ (vertical black dashed line) of a SNe~Ia sample with model parameters (here SALT2 stretch and colour) compatible with input model distributions within 3~$\sigma$. Indeed, up to this redshift limit the stretch distribution of selected SNe~Ia matchs almost perfectly the input distribution, while the colour distribution starts to diverge from the input distribution for a redshift $z > 0.05$. 

Furthermore, we observe that the KS test behaviour for low redshift (typically $z < 0.08$) is similar when comparing the parameter distributions after selection to the generated input distributions, as well as for true parameter (full lines) than for fitted parameter with SALT2 (dotted lines). This result reinforces the confidence in the fact that the volume limited sample as define before is not biased. We can also note that for higher redshitfs ($z > 0.08$) the KS test p-value increases for fitted parameters compared to generated ones. This effect can be interpreted by a spreading of stretch and colour distributions with fitted parameters due to the fact that SNe~Ia are fainter and then the uncertainty in the estimation of SALT2 parameters in the fitting of light-curves becomes more important. This spreading dilutes the distortion of the distributions after selection w.r.t. to the input ones.

This redshift dependency analysis of the input SALT2 parameters after selection explains the very small and the more consequent disagreement of the stretch and colour distributions, respectively, compared to the ZTF-DR2-Cosmo data integrated in the full redshift range ($z < 0.15$) as seen in Fig.~\ref{fig:sim_salt2}.

In conclusion, the redshift limit $z_{lim} \leq 0.06$ constitutes a robust estimation to define the volume limited sample of the DR2 dataset, i.e. a sample without any bias when the light-curves are fitted using the SALT2 model.

\section{Toward cosmological analysis}

The simulation framework was tested up to the Hubble-Lema\^{i}tre diagram. To this end, the distance modulus of the fitted SNe~Ia of the high-statistic simulation were computed by assuming the fiducial input parameters for the standardization: $\mu_i = m_{b,i} - (M_b - \alpha \, x_{1,i} + \beta \, c_i)$ for every ($i$) SN~Ia with $M_b = -19.3$, $\alpha = 0.14$ and $\beta = 3.15$.
Then the residuals to the fiducial input cosmology (Planck 2018~\citealt{2020A&A...641A...6P}) were deduced: $\Delta\mu_i = \mu_i - \mu(z_i)$, where $\mu(z_i)$ is the theoretical result of the fiducial cosmology for every SN~Ia redshift $z_i$. 

\begin{figure}
  \centering
  \includegraphics[width=1.0\linewidth]{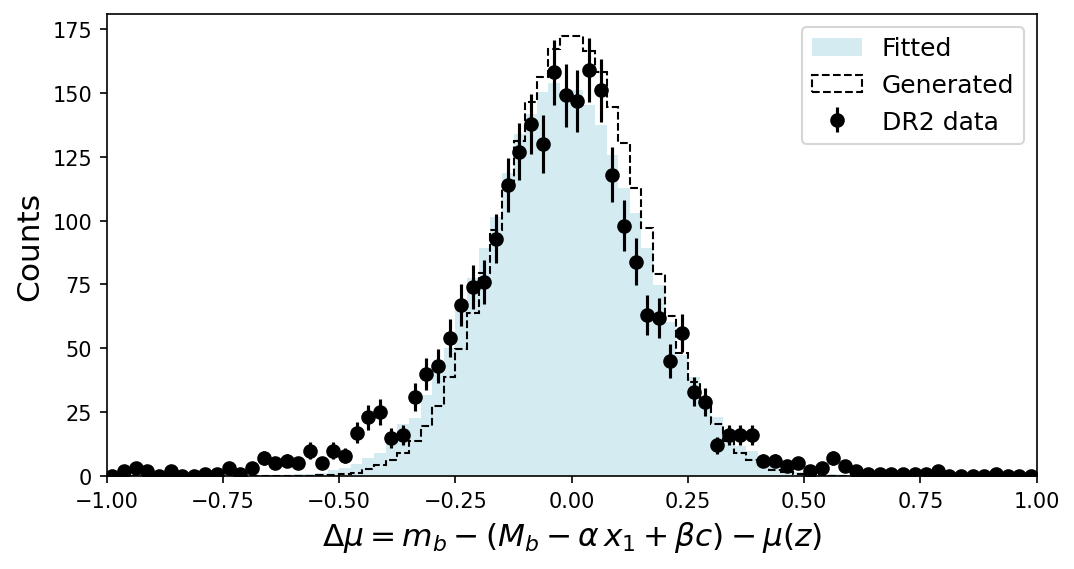}
  \caption{Hubble-Lema\^{\i}tre diagram residual distribution, assuming no environmental steps, for the DR2 (black points) and the simulation with fitted SALT2 parameters (full histogram) and the generated ones (open histogram). 
  } 
  \label{fig:sim_delta}
\end{figure}

The residual distribution is presented in Fig.~\ref{fig:sim_delta} with a comparison between DR2 data ($2624$ SNeIa) and simulation with fitted SALT2 parameters (full histogram) as well as generated parameters (open histogram). Firstly, the dispersion of the fitted residual distribution is close to the generated one, indicating that the residual is mainly driven by intrinsic dispersion ($\sigma_{\it int}$). Secondly, the fitted simulation reproduces overall the distribution measured from the DR2 dataset ($\chi^2_{/ndf}=2.00$), with small discrepancies that we will tackle in the upcoming DR2.5 cosmological data release. The robust agreement between our simulated samples and the DR2 dataset for the volume-limited sample, with $\chi^2_{/ndf}=1.81,1.61,1.26,1.23,1.63$ when considering the distribution of $z,m_B,x_1,c,\Delta\mu$ respectively, indicates that both the underlying distribution of parameters and selection of events are well understood, laying the foundations for a cosmological analysis. 

The DR2 dataset (and other literature samples) have observed significant correlations between the residual luminosity, standardisation parameters and measures of the local environment \citep[e.g.]{ginolin2024ztf_1,ginolin2024ztf_2}. To ensure that our estimates of the cosmological parameters are unbiased, these must be included in our simulation framework, which must also be updated to match the reprocessed `scene-model photometry`SMP \textcolor{red}{Lacroix et al.} and benchmarked against the widely used software, e.g. \texttt{SNANA}. This work will be released as part of the DR2.5 cosmological analysis, due for release in late 2025.


\section{Conclusion}

In this paper we presented the performance of the simulation framework -- \texttt{skysurvey} and its advanced version \texttt{skysurvey} -- developed for the Zwicky Transient Facility (ZTF) survey. The targeted simulation of the first phase of ZTF, from April 2018 and December 2020, was used to validated the full pipeline. The confrontation to the ZTF SN~Ia DR2 sample has shown that neither the science nor the difference images magnitude limit are able to reproduce the measured flux uncertainties. A tuning of the sky-noise with an increasing factor of 1.23, 1.17 and 1.20 for g, r and i-bands, respectively, has been derived to match flux uncertainties when using science image magnitude limit with an additional calibration error of 2.5\%, 3.5\% and 6\% for g, r and i-bands. The simulation using realistic SALT2 stretch and colour distribution has shown the ability of the \texttt{skysurvey} package to reproduce the ZTF SN~Ia DR2 sample. Finally, a redshift study of SALT2 parameters allowed to identify a volume limited sample, $z \leq 0.06$, i.e. an unbiased sample of about 1000 SNe~Ia. This volume limited sample is unique to perform new studies on SNe~Ia and especially analysis to improve their standardization procedure for future cosmological analysis. 

\begin{acknowledgements}
    Based on observations obtained with the Samuel Oschin Telescope 48-inch and the 60-inch Telescope at the Palomar Observatory as part of the Zwicky Transient Facility project. ZTF is supported by the National Science Foundation under Grant No. AST-1440341 and a collaboration including Caltech, IPAC, the Weizmann Institute of Science, the Oskar Klein Center at Stockholm University, the University of Maryland, the University of Washington, Deutsches Elektronen-Synchrotron and Humboldt University, Los Alamos National Laboratories, the TANGO Consortium of Taiwan, the University of Wisconsin at Milwaukee, and Lawrence Berkeley National Laboratories. Operations are conducted by COO, IPAC, and UW.
    The ZTF forced-photometry service was funded under the Heising-Simons Foundation grant \#12540303 (PI: Graham).
    SED Machine is based upon work supported by the National Science Foundation under Grant No. 1106171
    This work was supported by the GROWTH project \citep{Kasliwal2019pasp} funded by the National Science Foundation under Grant No 1545949.
    This project has received funding from the European Research Council (ERC)
    under the European Union's Horizon 2020 research and innovation programme
    (grant agreement n 759194 - USNAC). 
    P.R. acknowledges the support received from the Agence Nationale de la Recherche of the French government through the program ANR-21-CE31-0016-03.
    UB is supported by the H2020 European Research Council grant no. 758638
    GD is supported by the H2020 European Research Council grant no. 758638

    L.G. acknowledges financial support from AGAUR, CSIC, MCIN and AEI 10.13039/501100011033 under projects PID2020-115253GA-I00, PIE 20215AT016, CEX2020-001058-M, and 2021-SGR-01270.
    This work has been supported by the research project grant “Understanding the
    Dynamic Universe” funded by the Knut and Alice Wallenberg Foundation under Dnr KAW 2018.0067 and the {\em Vetenskapsr\aa det}, the Swedish Research Council, project 2020-03444
    LH is funded by the Irish Research Council under grant number GOIPG/2020/1387
    Y.-L.K. has received funding from the Science and Technology Facilities Council [grant number ST/V000713/1]
    KM is supported by the H2020 European Research Council grant no. 758638.
    T.E.M.B. acknowledges financial support from the Spanish Ministerio de Ciencia e Innovaci{\'o}n (MCIN) and the Agencia Estatal de Investigaci{\'o}n (AEI) 10.13039/501100011033 under the PID2020-115253GA-I00 HOSTFLOWS project, and from Centro Superior de Investigaciones Cient{\'i}ficas (CSIC) under the PIE project 20215AT016 and the program Unidad de Excelencia Mar{\'i}a de Maeztu CEX2020-001058-M.
    JHT is supported by the H2020 European Research Council grant no. 758638.

\end{acknowledgements}

\bibliographystyle{aa} 
\bibliography{ztfdr2_sim, ztfdr2_papers} 

\end{document}